\documentclass{appolb}
\usepackage[utf8]{inputenc}
\usepackage{bm}
\usepackage{bbold}
\usepackage{mathtools}
\usepackage[toc,page]{appendix}
\usepackage{dsfont}
\usepackage{combelow}
\usepackage[colorlinks=true,linkcolor=blue,citecolor=blue, urlcolor=blue]{hyperref} 
\usepackage{tikz}
\usepackage{graphicx}
\usepackage{subfigure}
\usepackage{scalerel,amssymb}
\DeclareMathAlphabet{\pazocal}{OMS}{zplm}{m}{n}

\newcommand{\xT}{\mathbf{x_\perp}}

\newcommand{\rT}{\mathbf{r_\perp}}

\newcommand{\kT}{\mathbf{k_\perp}}

\def\deltan{\hat\delta}

\begin{document}
\title{Tracing the emergence of collectivity phenomena in small systems%
\thanks{Presented at Quark Matter 2022 XXIXth International Conference on Ultra-relativistic Nucleus-Nucleus Collisions}%
}
\author{B.~Schenke
\address{Physics Department, Brookhaven National Laboratory, Bldg.\,510A, Upton, NY 11973, USA}
\\[2.5mm]
{S. Schlichting
\address{Fakultät für Physik, Universität Bielefeld, D-33615 Bielefeld, Germany}
}
\\[2.5mm]
\underline{P.~Singh}\thanks{Presenter}
\address{Department of Physics, P.O. Box 35, 40014 University of Jyv\"askyl\"a, Finland}
\address{Helsinki Institute of Physics, P.O. Box 64, FI-00014 University of Helsinki, Finland}
}
\maketitle
\begin{abstract}
We study initial state momentum correlations and event-by-event geometry in p+Pb collisions at $\sqrt{s}=5.02~\rm TeV$ by following the approach of extending the IP-Glasma model to 3D using JIMWLK rapidity evolution. On examining the non-trivial rapidity dependence of the observables, we find that the geometry is correlated over large rapidity intervals, while the initial state momentum correlations have a relatively short range in rapidity. Based on our results, we discuss implications for the relevance of both effects in explaining the origin of collective phenomena in small systems.
\end{abstract}

\section{Introduction}
Experimental evidence for collectivity in high multiplicity p+p/A \cite{Dusling:2015gta} collisions has lead to an increasing interest in understanding the origin of 
these long range azimuthal correlations in small systems. The general features of these correlations are similar to those observed in heavy-ion collisions where these structures are interpreted in terms of the system's response to initial geometry \cite{Gale:2013da}. For small systems, event geometry \cite{Schenke:2021mxx} and initial state correlations \cite{Lappi:2015vta} have been invoked as possible explanations of observed long-range rapidity correlations. In this proceeding, the relative contribution of both these mechanisms have been analyzed as a function of rapidity for pPb collision at $5.02$ TeV by using the $3+1$D IP-Glasma model that incorporates the longitudinal structure by including the JIMWLK evolution of the incoming nuclear distributions. The details of our work can be found in \cite{Schenke:2022mjv}.


\section{3D IP-Glasma model and event generation}
We follow the description \cite{Schenke:2016ksl} within the Color Glass Condensate (CGC) framework where the expectation value of an observable $O(y_{\rm obs})$ at any rapidity $y_{\rm obs}$ is calculated in terms of the high energy factorization of the projectile (p) and the target (Pb) \cite{Gelis:2008rw} as
\begin{align}
    O(y_{\rm obs})=O_{\rm cl}\Big(V_{\xT}^{p}(+y_{\rm obs}),V^{Pb}_{\xT}(-y_{\rm obs})\Big)
\end{align}
where $V^{p/Pb}_{x_\perp}$ denote the light like Wilson line at position $\xT$.

We generate a set of 4096 events from a set of proton and lead Wilson line configurations that are evolved to all rapidities of interest using JIMWLK evolution and collided at different impact parameters. 
For each event, we run a series of independent $2+1$~D classical Yang-Mills (CYM) simulations at intervals of $\Delta y=0.4$ in the $y_{\rm obs} \in [-2.4,+2.4]$ range, to get the rapidity dependence of observables 
according to the factorization scheme.

The centrality selection is based on the gluon multiplicity at mid-rapidity.
We will present results for two different sets of IR regulator used in the IP-Glasma model $(\tilde{m})$ and the JIMWLK equations $(m)$, namely $m=\tilde{m}=0.2\,{\rm GeV}$ and $m=\tilde{m}=0.8\,{\rm GeV}$.

\section{Results}

We first examine the longitudinal structure of the event geometry and the initial state momentum anisotropy by following the standard procedure in which the event geometry is characterized by eccentricities \begin{align}
    \varepsilon_n(y)=\frac{\int d^2\rT T^{\tau\tau}(y,\rT)~|\rT|^n e^{in\phi_{\rT}}}{\int d^2\rT T^{\tau\tau}(y,\rT)~|\rT|^n}\,
\end{align}
and the momentum anisotropy is given by the azimuthal anisotropy of the produced gluons 
\begin{align}
    v_2^{g}(y)=\frac{\int d^2\kT |\kT| \frac{dN_{g}}{dy d^2\kT}(y)e^{2i\phi_{\kT}}}{\int d^2\kT |\kT|~\frac{dN}{dy d^2\kT}(y)}\,.
\end{align} 

\begin{figure}%
    \includegraphics[width=0.32\textwidth]{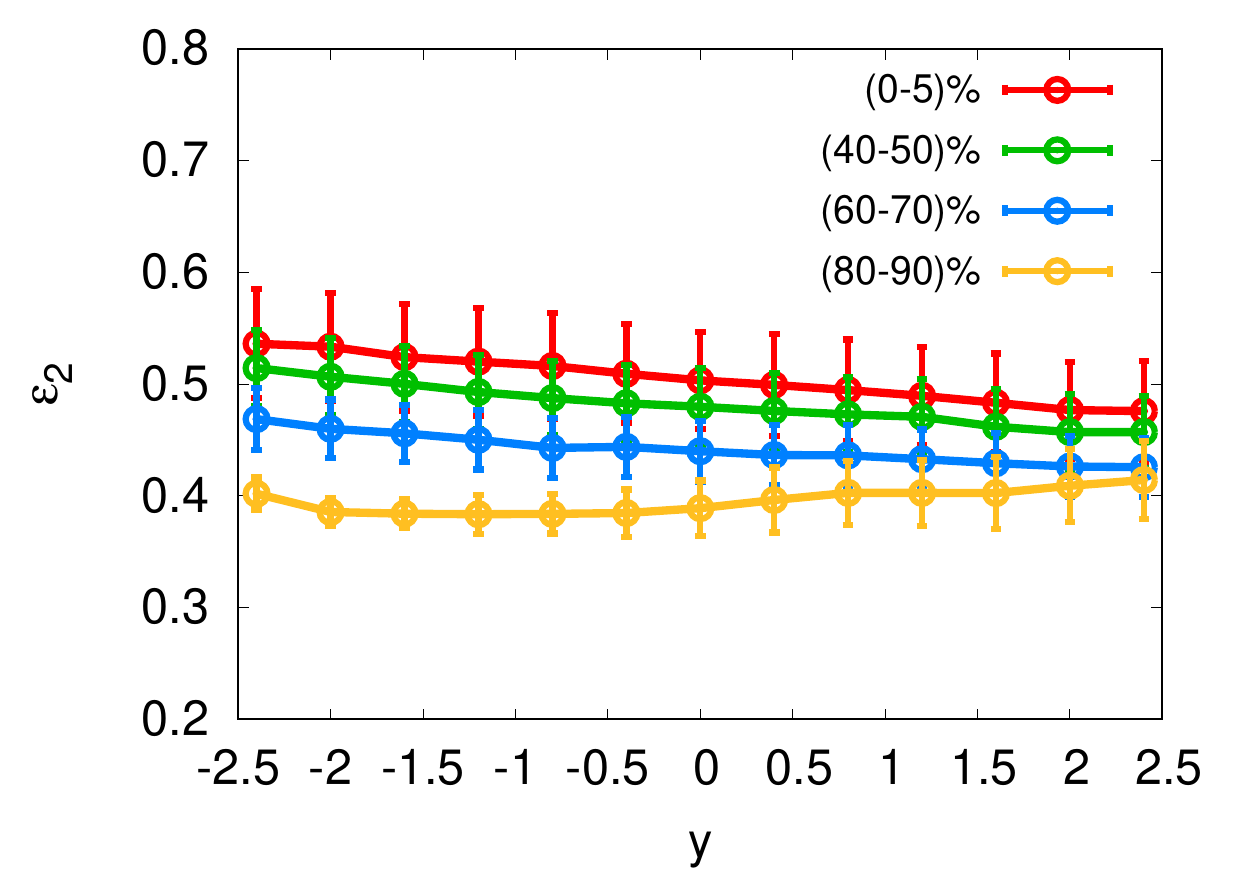}%
    \includegraphics[width=0.32\textwidth]{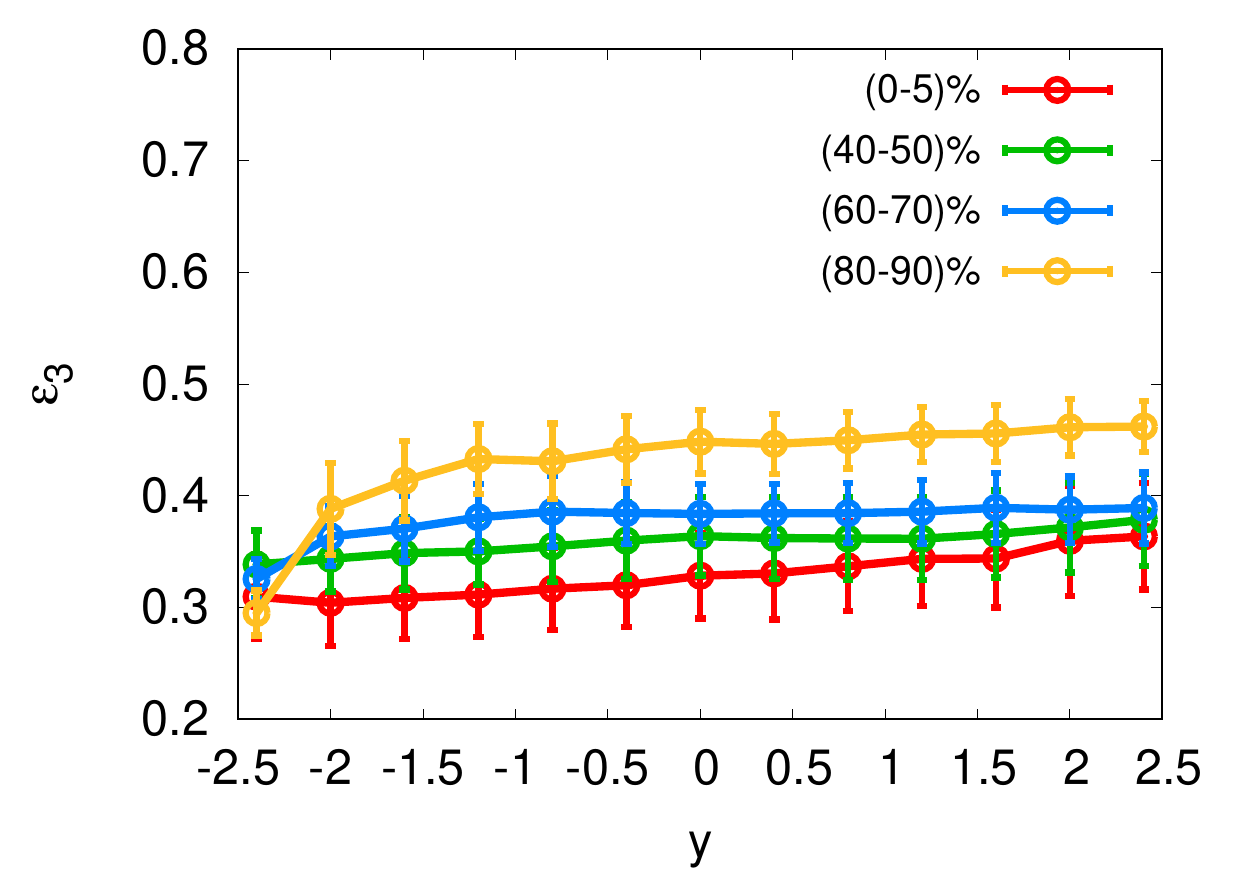}%
    \includegraphics[width=0.32\textwidth]{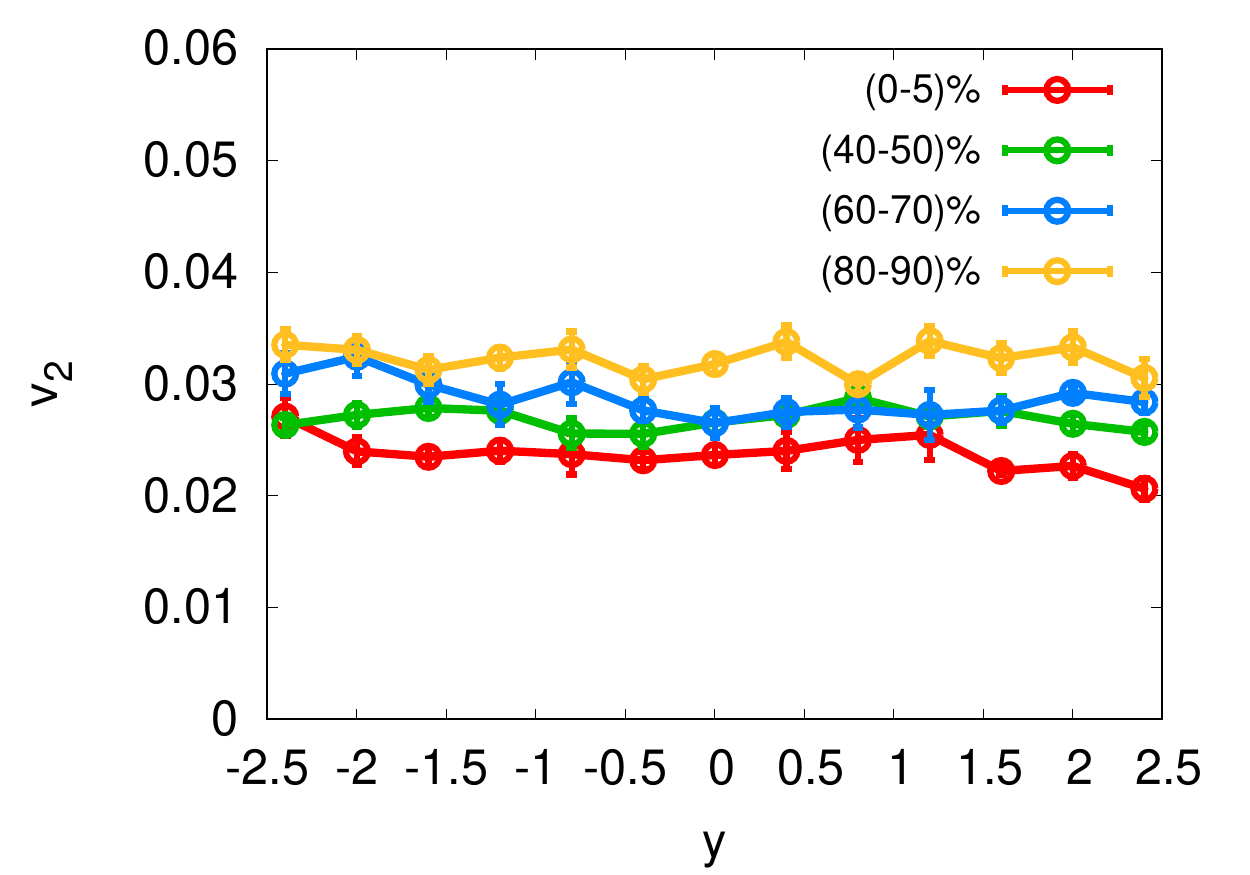}
    \caption{The rapidity dependence of $\sqrt{ \langle |\varepsilon_{2}(y)|^2\rangle}$ (left), $\sqrt{ \langle |\varepsilon_{3}(y)|^2\rangle}$ (middle) and $\sqrt{\langle|v_2^{g}(y)|^2\rangle}$ (right) for different centrality classes for $m=\tilde{m}=0.8~\rm GeV$}
    \label{fig:Ecc_and_v}
\end{figure}

Our results for the rapidity dependence of $\varepsilon_2(y)$ (left) and $\varepsilon_3(y)$ (middle), and  $v_2^{g}$ (right) are compactly summarised in Fig.~\ref{fig:Ecc_and_v}. We observe that $\varepsilon_2$ decreases, while the triangularity $\varepsilon_3$ increases with decreasing multiplicity.
The eccentricity $\varepsilon_2$ also decreases with increasing rapidity, while $\varepsilon_3$ shows a weaker rapidity dependence except for the most central and the most peripheral collisions, where it increases toward the Pb going direction. On the other hand, $v_2^g$ is largely independent of rapidity in all the centrality bins and grows monotonically with decreasing multiplicity.

We now analyze the rapidity decorrelation of geometry and initial state momentum correlations by computing the normalized rapidity correlation function $C^N_{\mathcal{O}}(y_1,y_2)=\left\langle {\rm Re}\big(\mathcal{O}(y_1)\mathcal{O}^{*}(y_2)\big) \right\rangle/\sqrt{\langle |\mathcal{O}(y_1)|^2\rangle\langle |\mathcal{O}(y_2)|^2\rangle}$ where
$\mathcal{O}$ is any of the previously defined observables.
\begin{figure}
    \centering
    \includegraphics[width=0.49\textwidth]{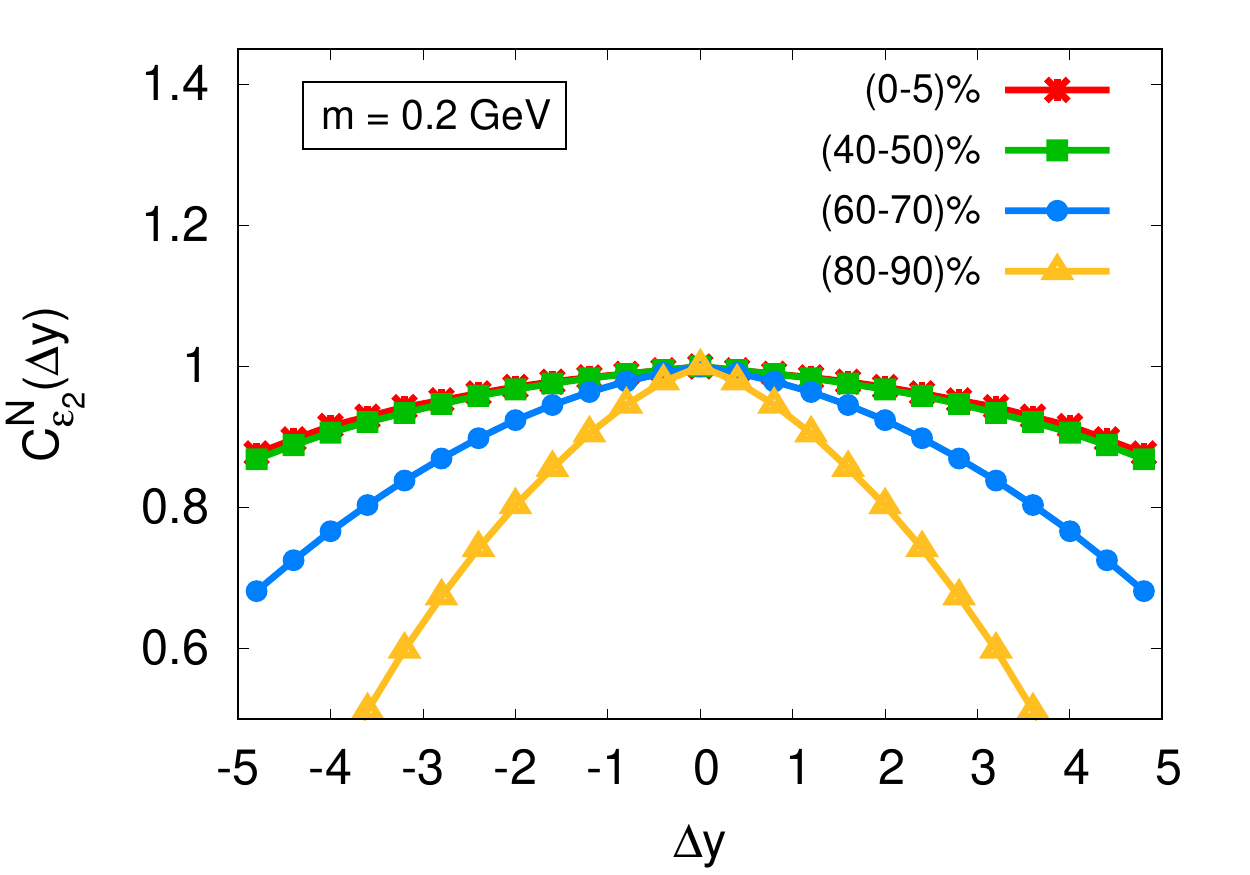}
    \includegraphics[width=0.49\textwidth]{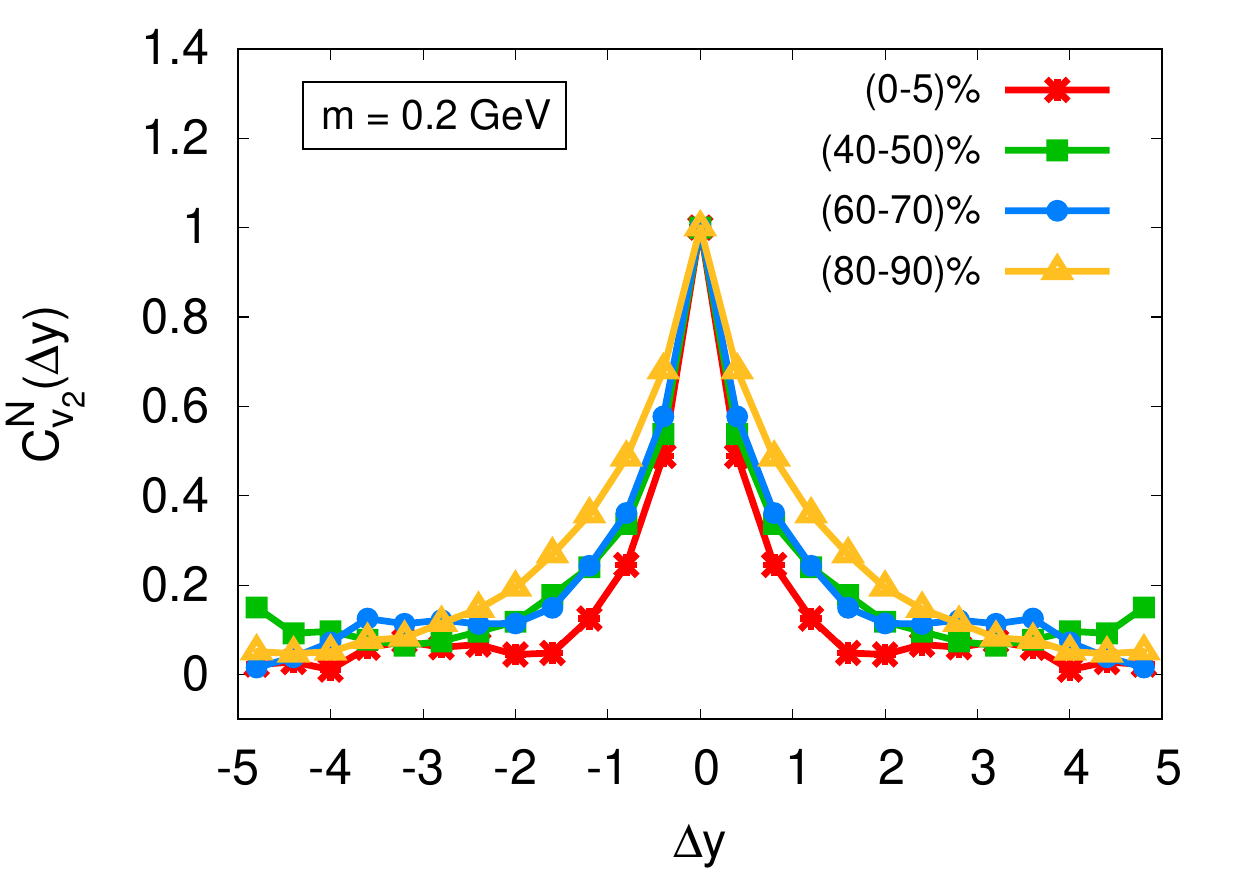}
    \caption{Normalized correlation function $C^{N}_{\mathcal{O}}(\Delta y)$ of eccentricity $\varepsilon_2$ (left) and initial state anisotropy $v_2$ (right) for different centrality classes using $m=\tilde{m}=0.2\,{\rm GeV}$ }
    \label{fig:CorrelationFunction}
\end{figure}
In Fig.\,\ref{fig:CorrelationFunction}, we show $C^N_{\varepsilon_{2}}(\Delta y)$ and $C^N_{v_2}(\Delta y)$ as functions of the rapidity difference $\Delta y$ for $m=\tilde{m}=0.2 \rm GeV$, which are obtained as $$C^{N}_{\mathcal{O}}(\Delta y)=\frac{1}{2y_{\rm max}-\Delta y}\int_{-y_{\rm max}+|\Delta y|/2}^{+y_{\rm max}-|\Delta y|/2} dY ~C^{N}_{\mathcal{O}}(Y + \Delta y/2,Y-\Delta y/2)\,.$$ The geometry decorrelates faster towards more peripheral events because it is easier to change the geometry of dilute events.
The decorrelation of the initial state momentum anisotropy shows the opposite centrality dependence, with the most peripheral collisions showing a slower decorrelation.
Comparing the results in the two panels, we see that the decorrelation of the initial state momentum anisotropy in $C^N_{v_2}(\Delta y)$ is much faster than the decorrelation of the event geometry in $C^N_{\varepsilon_{2}}(\Delta y)$. 


Finally, we study estimators for the correlation of mean transverse momentum $[p_T]$ and elliptical anisotropy $V_2$, which is defined as \cite{Bozek:2016yoj}
\begin{align}
     \hat{\rho}(V_2^2,[p_T])=\frac{\langle \deltan V_2^2 \,\deltan [p_T]\rangle}{\sqrt{\langle(\deltan V_2^2)^2\rangle\langle(\deltan [p_T])^2\rangle}}\,
\end{align}
The event-by-event fluctuations of an observable $O$ at fixed multiplicity are defined as $\deltan O \equiv \delta O-\big(\langle \delta O \delta N \rangle /\sigma_N^2\big)\delta N$ where $\delta O=O-\langle O\rangle$, $N$ is the multiplicity and $\sigma_N$ the variance of $N$ in a given centrality bin~\cite{Olszewski:2017vyg}. Since we are considering initial state quantities, we compute $\hat{\rho}$ by replacing $V_2$ with the initial state eccentricity $\varepsilon_2$ (or the initial momentum anisotropy $v_2$) and $[p_T]$ by the average entropy density $[s]=[e^{3/4}]$ where $e$ is the energy density, approximated as $T^{\tau\tau}$.
\begin{figure}
    \centering
    \includegraphics[width=0.49\textwidth]{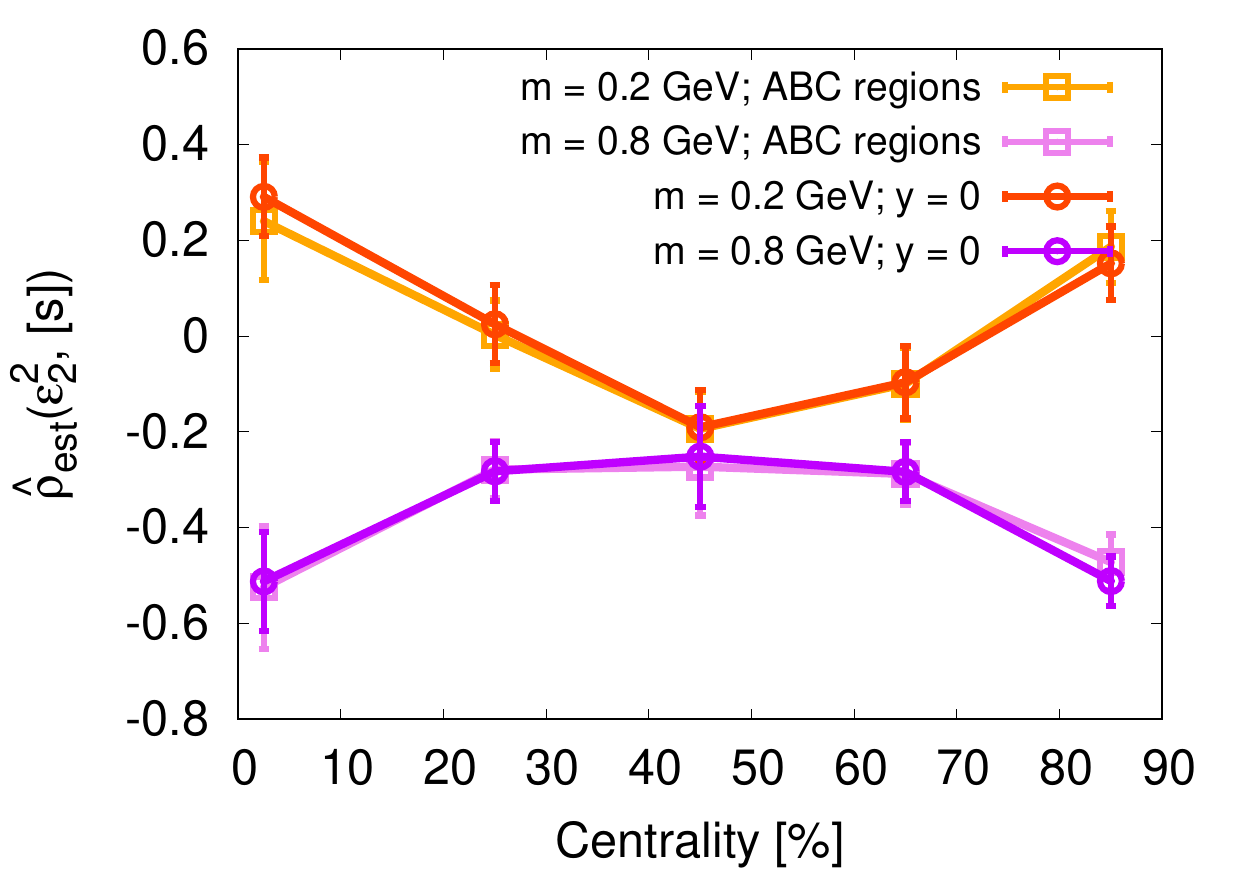}
    \includegraphics[width=0.49\textwidth]{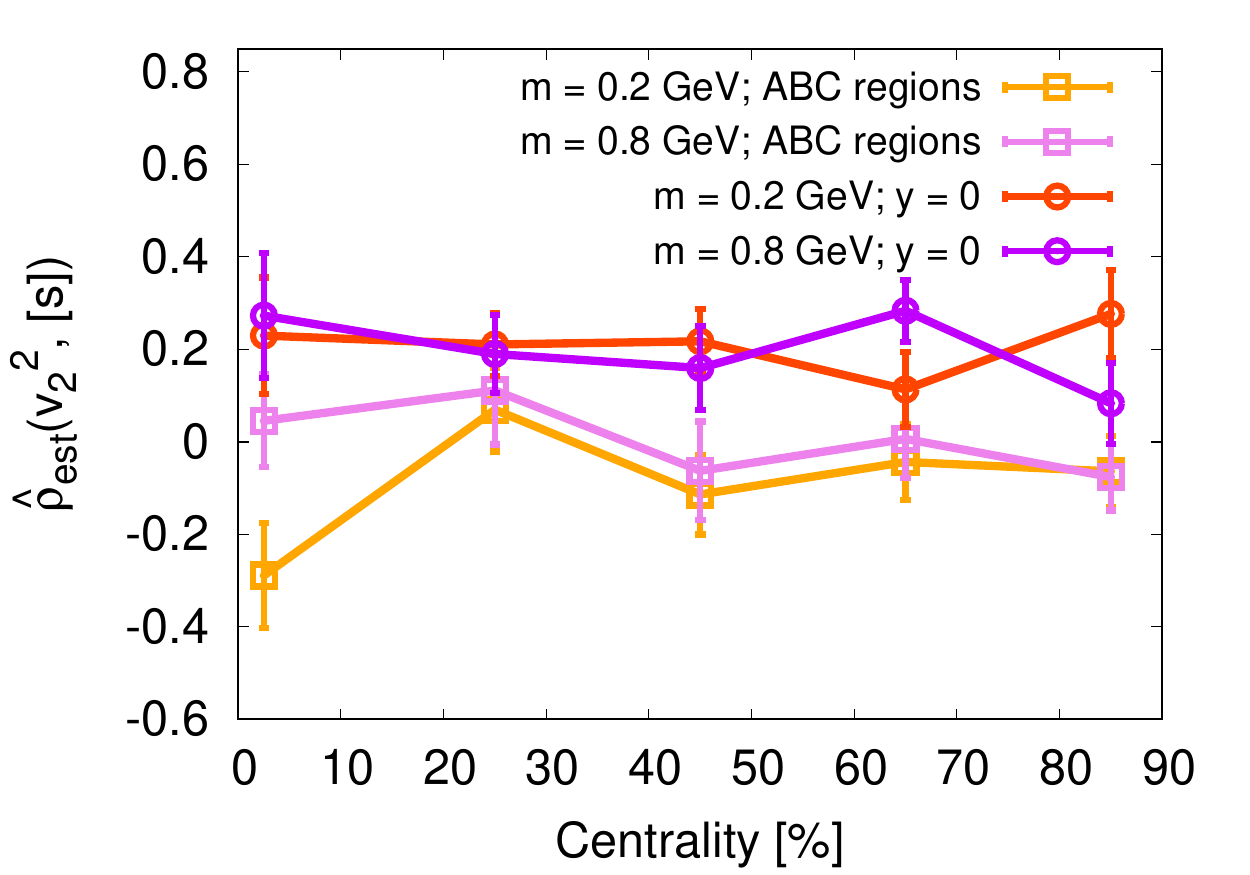}
    \caption{The estimators for geometry $\hat{\rho}_{\rm est}(\varepsilon_2^2,[s])$ (left) and initial state momentum anisotropy $\hat{\rho}_{\rm est}(\epsilon_p^2,[s])$ (right) as a function of centrality for two different values of $(m=\tilde{m})$. The $\hat{\rho}$ for ABC is obtained for different rapidity regions: region A with $-2.4<y<-0.8$, region B with $|y|<0.8$ and region C with $0.8<y<2.4$}
    \label{fig:rhoest}
\end{figure}

Our results for $\hat{\rho}_{\rm est}(\varepsilon_2^2,[s])$ (left) and $\hat{\rho}_{\rm est}(v_2^2,[s])$ (right) are summarized in Fig.~\ref{fig:rhoest}, where we present the measurement for two different bin selection methods. One uses all quantities at mid rapidity $y=0$ and the other uses three different rapidity bins (ABC regions) for the different components of $\hat{\rho}$. We find that for the larger $m=\tilde{m}$, the $\hat{\rho}_{\rm est}(\varepsilon_2^2,[s])$ is always negative, as can be expected from geometric considerations \cite{Giacalone:2020byk}. For $m=\tilde{m}=0.2\rm GeV$, we even find positive values for most central and most peripheral events which is in line with previous work \cite{Bozek:2020drh} where the geometric $\hat{\rho}$ correlator turned positive when increasing the system size. We also notice $\hat{\rho}_{\rm est}(\varepsilon_2^2,[s])$ is independent of the choice of the rapidity bins which is related to the weak decorrelation of the geometry observed. For $\hat{\rho}_{\rm est}(v_2^2,[s])$, we observe a positive correlation when quantities in estimators are taken at mid-rapidity, which is again in line with previous work \cite{Giacalone:2020byk}. However, for the ABC region $\hat{\rho}_{\rm est}(v_2^2,[s])$ is consistent with zero due to the rapid decorrelation of $v_2$ with rapidity.

\section{Conclusion \& Outlook}
We have presented results for rapidity dependent
quantities in p+Pb collisions, computed within the CGC framework. We computed the unequal rapidity correlations of both geometric and initial state momentum anisotropy, quantified by $\varepsilon_2$ and $v_2$, respectively, and observed that the geometry decorrelates much more slowly as a function of the rapidity difference, compared to the initial momentum anisotropy.
Beyond phenomenological applications of the 3-D IP Glasma model to collective flow in small and large systems, further theoretical progress towards the construction of a fully 3D
Wilson line configuration followed by 3+1D Yang-Mills
evolution, as explored in \cite{Schlichting:2020wrv,Ipp:2021lwz}, will be desirable in the future.

\textit{Acknowledgments:} We acknowledge support by the U.S. Department of Energy under Contract No. DE-SC0012704 (B.S) and Deutsche
Forschungsgemeinschaft through the CRC-TR 211 project number 315477589 (S.S and P.S). 
This research used resources of NERSC, which is supported by the Office of Science of the U.S. DOE under Contract No. DE-AC02-05CH11231.
\bibliographystyle{elsarticle-num}
\bibliography{bib}
\end{document}